\documentclass[lettersize]{IEEEtran}
\usepackage[utf8]{inputenc}
\usepackage{amsmath,amsfonts}
\usepackage{algorithmic}
\usepackage{array}
\usepackage[caption=false,font=normalsize,labelfont=sf,textfont=sf]{subfig}
\usepackage{textcomp}
\usepackage{stfloats}
\usepackage{url}
\usepackage{verbatim}
\usepackage{graphicx}
\def\BibTeX{{\rm B\kern-.05em{\sc i\kern-.025em b}\kern-.08em
    T\kern-.1667em\lower.7ex\hbox{E}\kern-.125emX}}
\usepackage{balance}
\usepackage{xspace}
\newcommand{\Lcano}{\ensuremath{L_{\text{C}}}\xspace}
\newcommand{\Lleft}{\ensuremath{L_{\text{L}}}\xspace}
\newcommand{\Lright}{\ensuremath{L_{\text{R}}}\xspace}
\newcommand{\Lsdi}{\ensuremath{L_{\text{SDI}}}\xspace}
\newcommand{\Lsd}{\ensuremath{L_{\text{SD}}}\xspace}

\newcommand{\SEG}{\ensuremath{\mathcal{A}}\xspace}
\usepackage{tipa}
\newcommand{\ipaW}{\textipa{w}}
\newcommand{\ipaIH}{\textipa{\i}}
\newcommand{\ipaAH}{\textipa{\textturnv}}

\usepackage[
    style=ieee,
    doi=false,
    isbn=false,
    url=false,
    eprint=false,
    date=year, 
]{biblatex}
\usepackage{xcolor}

\begin{document}
\title{Segmentation-Free Goodness of Pronunciation}
\author{Xinwei Cao, Zijian Fan, Torbjørn Svendsen, ~\IEEEmembership{Senior Member,~IEEE}, Giampiero Salvi, ~\IEEEmembership{Senior Member,~IEEE}
\thanks{All authors are with the Department of Electronic Systems, Norwegian University of Science and Technology (NTNU), Norway. \{xinwei.cao, zijian.fan, torbjorn.svendsen, giampiero.salvi\}@ntnu.no}
\thanks{This research was partly funded by the NordForsk Teflon project (\#103893) and by the Research Council of Norway SCRIBE project (\#322964).}
}

\markboth{Journal of \LaTeX\ Class Files,~Vol.~18, No.~9, September~2020}%
{How to Use the IEEEtran \LaTeX \ Templates}

\maketitle




\begin{abstract} 
Mispronunciation detection and diagnosis~(MDD) is a significant part in modern computer-aided language learning~(CALL) systems.
Most systems implementing phoneme-level MDD through goodness of pronunciation (GOP), however, rely on pre-segmentation of speech into phonetic units.
This limits the accuracy of these methods and the possibility to use modern CTC-based acoustic models for their evaluation.
In this study, we first propose self-alignment GOP (GOP-SA) that enables the use of CTC-trained ASR models for MDD.
Next, we define a more general segmentation-free method that takes all possible segmentations of the canonical transcription into account (GOP-SF).
We give a theoretical account of our definition of GOP-SF, an implementation that solves potential numerical issues as well as a proper normalization which allows the use of acoustic models with different peakiness over time.
We provide extensive experimental results on the CMU~Kids and speechocean762 datasets comparing the different definitions of our methods, estimating the dependency of GOP-SF on the peakiness of the acoustic models and on the amount of context around the target phoneme.
Finally, we compare our methods
with recent studies over the speechocean762 data
showing that the feature vectors derived from the proposed method achieve state-of-the-art results on phoneme-level pronunciation assessment.

\end{abstract}


\section{Introduction}
\IEEEPARstart{C}{omputer-aided} language learning (CALL) and computer-assisted pronunciation training~(CAPT) are becoming more important and helpful among language learners and teachers both because they are ubiquitously available and because they maintain a high degree of self-controlled manner of study. 
One of the desirable features for these systems is the ability to provide instant feedback and intervention when the learner makes any mispronunciation.
However, the automatic mispronunciation detection and/or diagnosis (MDD) modules currently available are not sufficiently reliable.

MDD can be performed at different linguistic levels: phoneme, word, and utterance.
One of the main challenges with MDD is the scarcity of data that is specifically annotated for the task, including information about mispronunciations.
This problem is especially severe for phoneme-level assessment which is the focus of this paper.

Witt et al.~\cite{witt2000phone} proposed a widely used method for MDD at the phoneme level that is based on a measure called goodness of pronunciation~(GOP).
The advantage of this method is that it relies on acoustic models exclusively trained for automatic speech recognition (ASR).
The ASR models are used to score new utterances, and only a small amount of MDD annotations are required to optimize a detection threshold that separates acceptable pronunciations from mispronunciations.
ASR models can also be used to transcribe speech at the word, character, or phoneme level and detect mispronunciations by comparing the resulting output sequence with the canonical pronunciation or orthographic transcription, respectively~\cite{cnn-rnn-ctc, fu2021textdependent, textdep2, hybrid, getman2022wav2vec2}.

Another approach is to train end-to-end models for MDD.
However, training such models from scratch would require large amounts of MDD annotated data.
A solution is to use foundation models that are either trained on large amounts of unlabeled speech data, or fine-tuned for ASR.
These models can be further fine-tuned with small amounts of MDD annotated data for the MDD task.
For example, Xu et al.~\cite{xu2021explore} follow this strategy starting from a Wav2Vec2.0 model~\cite{w2v2}
whereas, the authors of~\cite{ssl1, ssl2} start from a HuBERT model~\cite{hubert}.
Liu et al.~\cite{Liu2023ZeroShotAP} attempt to perform MDD at utterance-level based on hidden representations directly derived from the foundation models without further fine-tuning.



In this paper, our goal is to combine the advantages of the GOP method, the foundation models, and end-to-end ASR training based on CTC loss~\cite{CTC}.
This raises a number of challenges due to the fact that GOP requires segmentation of speech into phonetic units.
This is typically obtained by forced alignment of the canonical pronunciation with the spoken utterance. 
If the pronunciation is correct, the obtained phonetic boundaries may vary due to coarticulation effects.
In case of mispronunciations, the segmentation may be even less reliable.
Finally, CTC trained models tend to give activations that are not necessarily aligned with acoustic speech segmentation.
In \cite{cao24b_interspeech} we introduced a framework for combining CTC trained models and GOP.
In particular, we introduced 1) a self-aligned GOP method that uses the same activations from CTC trained models for alignment and GOP evaluation, and 2) a segmentation-free GOP method that can assess a specific speech segment without committing to a particular segmentation.
In this paper, we enhance these methods by making the following novel contributions:
\begin{itemize}
    \item We define proper segment length normalization for our segmentation-free GOP definition without committing to a particular segmentation.
    \item We provide a theoretical derivation of the segmentation-free GOP method that exposes the assumptions required for its definition.
    \item We introduce a novel implementation of our method that eliminates numerical problems.
    \item We provide extensive experimental results that compare the different methods we propose to the state-of-the-art on binary and ternary MDD tasks.
    \item We provide an analysis of the effect of peakiness of the model on the segmentation-free GOP method.
    \item We provide an analysis of the effect of context length (utterance length) on the segmentation-free GOP method.
    \item We make all code available.\footnote{https://github.com/frank613/CTC-based-GOP}
\end{itemize}

The paper is organized as follows: in Section~\ref{sec:Backgrounds} we provide the background and review the related work that is necessary to understand our methods.
In Section~\ref{sec:Methods} we give our GOP definitions and theoretical derivations showing the assumptions that are implicit in them.
Sections~\ref{sec:experiments} and \ref{sec:results} report on our experimental settings and empirical results.
Finally, Section~\ref{sec:conclusions} concludes the paper. 
\begin{figure}[t]
  \centering
  \includegraphics[width=\columnwidth]{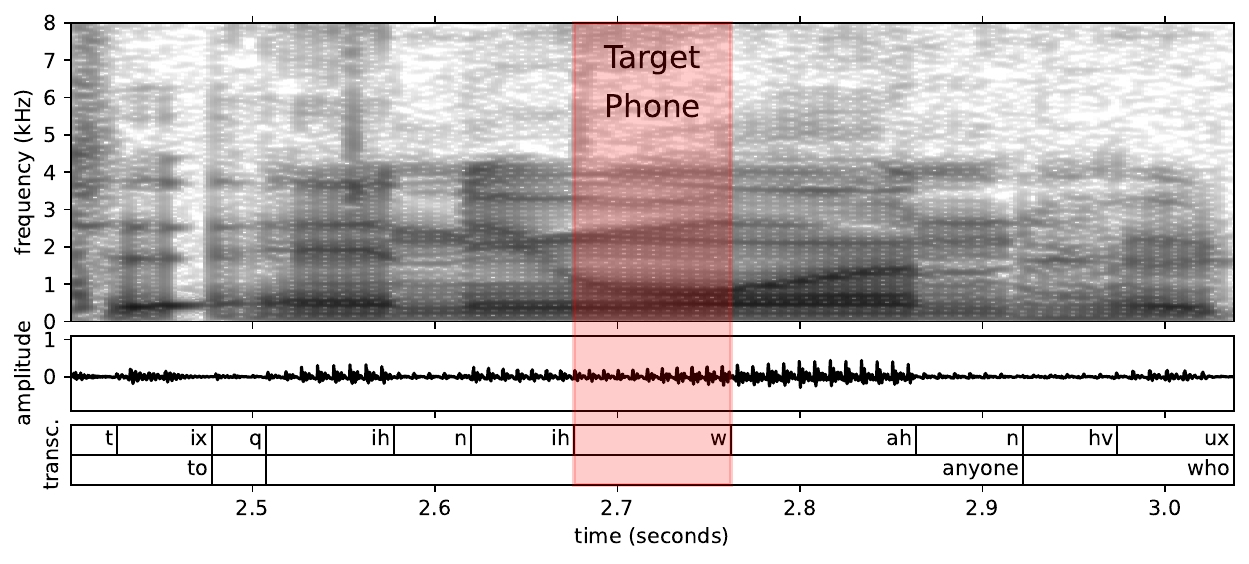}
  \caption{Standard GOP methods use forced alignment to segment speech for analysis.}
  \label{fig:gop-goal}
\end{figure} 

\section{Background and Related Work}
\label{sec:Backgrounds}
In this section we give background information and review the related work that are necessary to understand the proposed method.
We will focus on phoneme-level pronunciation assessment that is the goal of this paper.
We first need to distinguish between continuous measures for pronunciation assessment and the specific task of MDD.
In the first case, the goal is to introduce a metric that can indicate how far a specific pronunciation of a phoneme is from the acceptable variability in the language.
An example of this kind of measure is GOP.
The MDD task, on the other hand, is to provide a binary or, sometimes ternary, decision on whether the pronunciation is correct or not.
This task can be performed 1) on a single GOP-like score by defining thresholds for the different output classes, 2) by more advanced classifiers based on multidimensional engineered features, or 3) by end-to-end systems that take raw speech as input.

In this paper, we focus on MDD methods that are based on GOP scores and GOP features.
We will first present the different definitions of GOP in Section~\ref{sec:gop_definitions},
the problem arising from the need for accurate speech segmentation in Section~\ref{sec:alignment_issues}, and, finally, the problem of using end-to-end peaky models with GOP in Section~\ref{sec:peaky_behavior}.

\subsection{The definition of GOP}
\label{sec:gop_definitions}
GOP was initially proposed as a measure of how closely the pronunciation of a specific phoneme matches its expected canonical pronunciation~\cite{witt2000phone}.
This is a segmental measure that requires time-aligned phonetic annotations (see, e.g., Figure~\ref{fig:gop-goal}).
Witt's original definition of GOP~\cite{witt2000phone} corresponds to the log posterior of the canonical phoneme $l_i$ given the sequence of observations $O_{t_1}^{t_2} = \{o_{t_1}, \dots, o_{t_2}\}$ in the segment under test, normalized by the sequence length:
\begin{equation}
    \label{eqa:gop}
    \text{GOP}(l_i)=\frac{\log(\mathit{p}(l_i|O_{t_1}^{t_2}))}{t_2-t_1}.
\end{equation}
Mispronunciations are detected on the basis of the GOP value and an empirically determined threshold. 

The estimation of the posteriors was originally implemented using Hidden Markov Models and Gaussian Mixture Models~(HMM-GMM) and by using Bayes rule:
\begin{equation}
    \label{eqa:gop2}
    \textit{p}(l_i|O_{t_1}^{t_2})=\frac{\mathit{p}(O_{t_1}^{t_2}|l_i)\mathit{P}(l_i)}{\sum_{q\in Q} \mathit{p}(O_{t_1}^{t_2}|q)\mathit{P}(q)},
\end{equation}
where $\mathit{P}(q)$ represents the prior probability of each phoneme in the phoneme inventory $Q$ and the likelihood $\mathit{p}(O_{t_1}^{t_2}|l_i)$ can be directly evaluated using the acoustic model, for example with the forward algorithm.
For efficiency reasons, Witt approximates the summation in the denominator by obtaining the best path over the full utterance through a phone-loop model.
From Eq.~\eqref{eqa:gop2}, GOP is then the ratio of the likelihood of the segment estimated from the canonical phone model (numerator) and the likelihood of the best path (allowing any phone sequence) through the segment (denominator). 

The phoneme boundaries $t_1$ and $t_2$ could be obtained by human annotations.
However, this is not practical:
During model training, annotating large amounts of data would be too costly.
More importantly, this would make the methods not suitable for providing immediate and automated feedback to the students.
The segmentation step is, therefore, commonly automated by forced alignment using the canonical pronunciation  $\Lcano = \{l_1, \dots, l_{|\Lcano|}\}$ and a phonetic model trained for ASR.
This model does not need to be the same as that used for the GOP evaluation, and it is typically a context-independent HMM-GMM model.

Although Witt's GOP has been successful, many later works argue that the phone-loop approximation is not reliable in estimating the denominator of Eq.~\ref{eqa:gop2}.
For example, Lattice-based GOP~\cite{Song2010LatticebasedGI} includes contributions from N-best hypotheses to avoid underestimating the value of the denominator.
In~\cite{luo2009analysis}, the authors show performance improvements that can be obtained when the phone-loop is evaluated multiple times over the target segment for each phoneme  rather than once along the whole utterance.
This implementation follows more closely the original definition of the denominator in Eq.\ref{eqa:gop2}.

Phonological rules from the learner's first language~(L1) can also be included in the GOP methods, where the acoustic models are trained with target language~(L2) only, to achieve better accuracy~\cite{ERN, DUDY201862}.

The models used for estimating the GOP values are trained for ASR and therefore do not need any data that are annotated specifically for the MDD task.
However, in order to detect mispronunciations, 
a small amount of human-annotated data is required to estimate the optimal threshold that separates good pronunciations from bad ones.
These thresholds are usually phoneme-dependent but even finer-grained definitions (for example speaker-dependent) may improve accuracy (e.g. \cite{kanters09_slate}).

Using hybrid HMM-Deep Neural Networks~(DNN) acoustic models has been shown to improve MDD performance~\cite{hu13_interspeech}.
In this case, the posterior defined in Eq.~\ref{eqa:gop} can be evaluated directly from the posterior probabilities $p(l_i|o_t)$ from the softmax output of the last DNN layer for the acoustic frames in the target phoneme.
The corresponding GOP definition is then the average of the log posterior probabilities:
\begin{equation}
    \text{GOP-Avg}(l_i) = \frac{1}{t_2-t_1}\sum_{t=t_1}^{t_2}\log\left({\mathit{p}(l_i|o_t)} \right).
    \label{eqa:gop-dnn}
\end{equation}

The DNN is often trained frame-wise with cross entropy~(CE) loss.
Similarly to the HMM-GMM based GOP, the segment boundaries $t_1$ and $t_2$ are obtained from an external aligner that is also used to provide the frame-wise labels for training the DNN.
We call this method external alignment (EA) in the following.
In~\cite{improved_DNN}, the authors show that the performance of DNN-based GOP can be further improved by merging the probabilities of fine-grained sub-phoneme symbols~(senones) to phonemes.\footnote{To compare with CTC-based methods (GOP-CTC-SF and GOP-CTC-SA) where GOP is also operated at phoneme level, our later discussion and experiments are all based on models trained directly at phoneme level.}

In Eq.~\ref{eqa:gop-dnn} each frame is evaluated independently using the output of the DNN, neglecting the transition probabilities in the HMM.
In~\cite{HMM_tran_GOP}, the author argues that these probabilities can contribute with contextual information and including them in the GOP evaluation can improve performance.
In~\cite{context-aware-GOP}, transition and duration factors are introduced to increase context awareness in GOP.

One of the issues with using GOP for MDD is that the definitions in Eqs.~\ref{eqa:gop2} and~\ref{eqa:gop-dnn} only account for substitution errors.
However, deletion and insertion of phonemes could also be potential sources of mispronunciations.
This is one of the limitations that we address in this work.

\subsection{The alignment issues for GOP}
\label{sec:alignment_issues}
There are several issues with traditional definitions of GOP related to speech segmentation.
The first is the ability to perform a perfect alignment of phonetic segments to the recorded speech.
Phonetic segmentation is an ill-posed problem because it is based on the assumption that speech is produced as a sequence of discrete events.
However, coarticulation effects in speech production question the existence of explicit phonetic boundaries.
This explains the disagreement on segmentation even between trained phoneticians.
An example is shown in Figure~\ref{fig:gop-goal}, where it is difficult to tell the exact beginning and end of the sonorant [\ipaW] as it happens in the transition of two vowel sounds [\ipaIH] and [\ipaAH].

Even assuming the existence of well-defined phonetic boundaries, alignment errors may occur in case of mispronunciations because the models used for segmentation are usually trained on correctly pronounced speech.
Furthermore, as studied in~\cite{the_impact_of_FA, perform_FA, L2-GOP}, forced alignment may be unreliable due to speaker variability: age, accent, dialect, health condition, to name a few.
The uncertainty over phonetic boundaries has a strong impact on traditional GOP definitions which consider the boundaries as deterministic.
%
%

Another important limitation of traditional GOP emerges from the characteristics of the models and loss functions used in modern ASR training.
Even assuming perfect segmentation of speech, there is no guarantee that the activations of the models used for GOP evaluation are aligned with this segmentation.
A typical case is with end-to-end transformer-based models where the alignment between input speech and output symbols is somewhat arbitrary.
This aspect is rarely emphasized in the literature, where the method used for alignment is often not specified.

In \cite{GPP} the authors propose to use the general posterior probability (GPP) to mitigate alignment issues by allowing evaluation over any segments that overlap with the target segment.
Zhang et al.~\cite{textdep3} propose a pure end-to-end method that uses a sequence-to-sequence model without having to segment the speech.
However, this method requires a large number of human annotations for MDD and an additional step that compares the canonical phoneme sequence with the human-annotated sequence before training the model.

In this work, we propose two methods to relax the dependency of GOP on the accuracy of segmentation.
The first has the goal of reducing the effect of misalignment between the acoustic model activations and the speech segments, the second is completely segmentation-free.
The latter also allows for insertion and deletion errors that were mentioned in Section~\ref{sec:gop_definitions}.

\begin{figure*}[t]
  \centering
  \includegraphics[width=\linewidth]{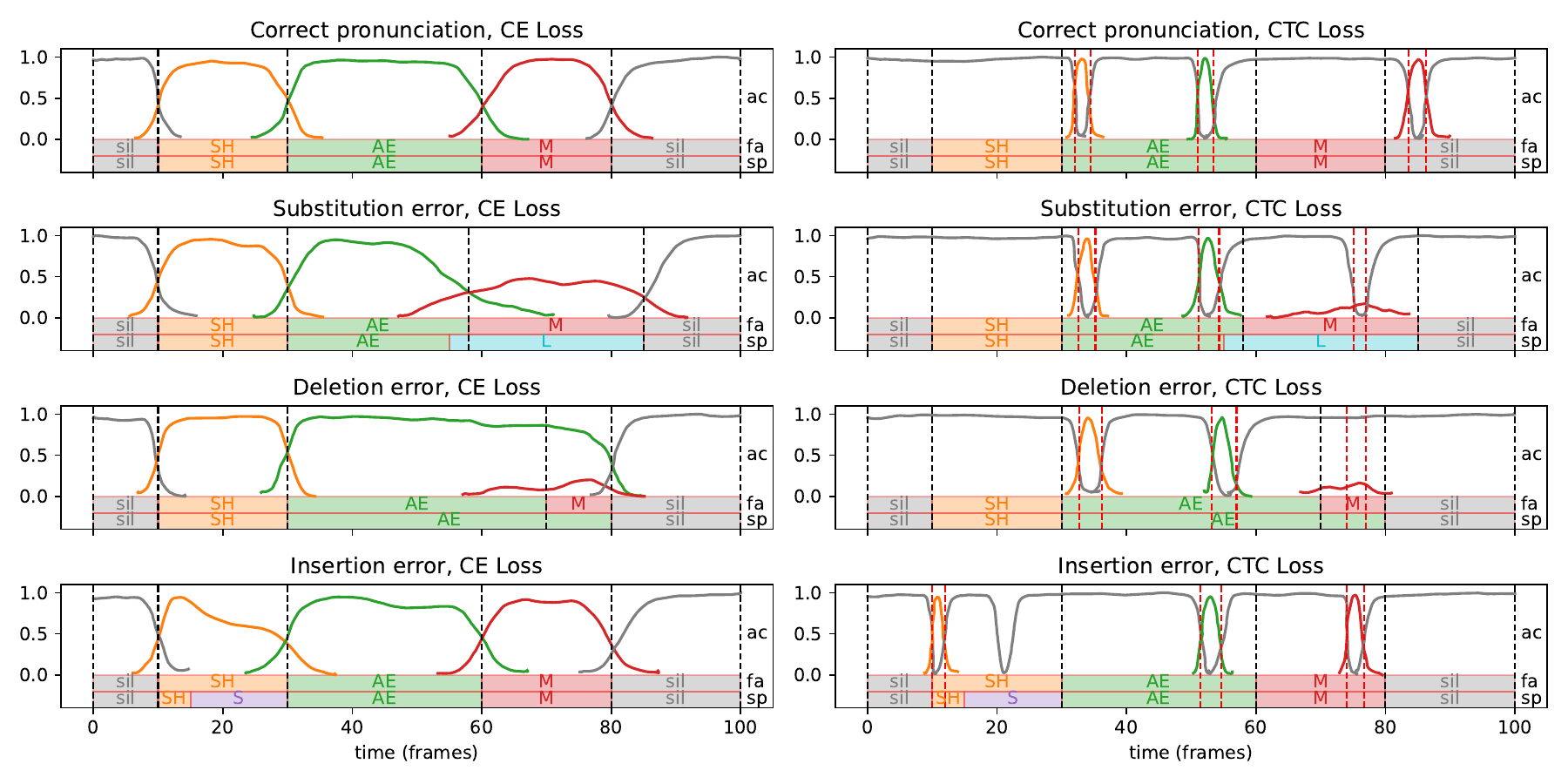}
  \caption{Illustration of the issues with standard GOP methods for models trained with CE loss (left) and CTC loss (right). Each plot corresponds to a different kind of mispronunciation and is divided in three parts: ``sp'' shows that segments that were actually spoken; ``fa'' shows the forced alignment based on the canonical pronunciation; ``ac'' shows the activations from the model and for the canonical segments. Finally, for the CTC models, the red dashed lines show the alternative forced alignment from the GOP-CTC-align method. }
  \label{fig:illustration-alignment-problems}
\end{figure*}  

\subsection{End-to-end ASR models, CTC and peaky behavior}
\label{sec:peaky_behavior}
In this section, we introduce some details of the Connectionist Temporal Classification (CTC) loss used in modern end-to-end ASR models.
This premise is important to understand our methods in Section~\ref{sec:Methods}.

End-to-end ASR models were initially introduced to map speech to output symbols, typically characters\footnote{The methods for MDD described in this paper use phonemes as output symbols. However, all the arguments in this section apply regardless of the nature of the output symbols.}, directly.
In general, the sequence of output symbols $L = \{l_1, \dots, l_{|L|}\}$ has a different rate that the input speech feature vectors $O_1^T = \{o_1, \dots, o_T\}$.
To overcome this problem in training, the CTC loss~\cite{CTC} was introduced.
This makes use of an additional blank ($\phi$) output symbol.
Given a speech sequence $O_1^T$ of length $T$, we define $V$ as the set of all output symbols including $\phi$.
Then any vector $u = \{u_1, \dots, u_T\} \in U = V^T$ is an alignment path between the input sequence and the output symbols.
The probability of a path $u$ under the assumption of conditional independence is $p(u| O_{1}^{T})=\prod_{t=1}^{T}p(u_t|o_t)$.
As a consequence $\sum_{u \in U}p(u| O_{1}^{T}) = 1$ representing all possible paths given the model and the speech.
During training, for a given target output sequence $L$ and the speech sequence $O_{1}^{T}$, the model learns to minimize the following loss:
\begin{equation}
   \label{eqa:CTC-loss}
    \mathcal{L}_{\text{CTC}}(L) = -\log\sum_{\pi \in {\{u:\mathcal{B}(u) = L\}}} p(\pi|O_1^T) 
\end{equation}
where $\pi$ spans over the paths that can be mapped to $L$ using the many-to-one function: $\mathcal{B}: U \rightarrow L$ by removing all repeated symbols and followed by removal of the blank symbols, e.g., $\mathcal{B}(a{\phi}b{\phi}b{\phi}cc{\phi})$ $=$ $abbc$. 

Since training the CTC does not require frame-level alignment, the timing information tends to be ignored by the model.
A well-known phenomenon of CTC-trained model is the ``peaky'' behavior of model output, as illustrated in the right column of Figure~\ref{fig:illustration-alignment-problems}.
There are two dimensions of peakiness: ``peaky over time'' (POT) and ``peaky over state'' (POS).
POT behavior corresponds to the fact that the blank symbol is activated for most of the time steps, whilst the non-blank symbols that correspond to the target label sequence only become activated for a few time steps. 
Forced alignment (Viterbi search) performed with CTC-trained models can result in distorted alignment between input speech and output symbols.
On the other hand, POS behavior is observed because the model activations at each frame are always close to 1.0 for one output and 0.0 for all the others.

For these reasons, CTC is not feasible for applications where the time alignment is essential, such as voice activity detection, speech segmentation, or conventional GOP evaluation for MDD.
An indication of this is the lack of literature in these areas that use CTC-based models.
However, several works tried to mitigate the peaky behavior of CTC-based models for ASR, e.g.~\cite{why_peaky, CTC-ME, huang2024peakyaccuratectcforced}.

The two methods proposed in this paper, have as goal to make it possible to use CTC-trained models with phoneme output targets as a basis for computing GOP for MDD tasks.

\section{Methods}
\label{sec:Methods}
The goals of our methods are (i)~to make it possible to use CTC-based models for a reliable evaluation of GOP, (ii)~to reduce the sensitivity of GOP to precise speech segmentation and to consider the uncertainty of phonetic alignment, (iii)~to introduce context awareness into the definition of GOP and (iv)~to extend the method of GOP to allow for detecting insertion and deletion errors as well as substitutions.

To achieve these goals we propose two methods that will be detailed in Section~\ref{sec:gop-self-segmentation} and \ref{sec:gop-segmentation-free}.

\subsection{Self-alignment GOP (GOP-SA)}
\label{sec:gop-self-segmentation}
Firstly, we consider the problem of mismatch between the segmentation of speech into phonetic units and the activations of the models used for GOP estimation as mentioned in Section~\ref{sec:alignment_issues}.
This phenomenon is exemplified in Figure~\ref{fig:illustration-alignment-problems} and is especially critical for CTC trained models (right plots) which exhibit peaky activations for each phoneme that may not correspond in time with the corresponding speech segment.
In order to address the problem, we propose to use the same GOP definition as for DNN models (Eq.~\ref{eqa:gop-dnn}) but to perform the alignment of the target segment ($t_1,t_2$) based on the same model used for GOP evaluation instead of an external forced aligner.
We refer to this method as self-alignment GOP (GOP-SA).
Figure~\ref{fig:illustration-alignment-problems} shows this alignment with red dashed vertical lines for the CTC-Loss trained models.
We want to stress that the goal of alignment in this method is not to find the segment corresponding to the target phone, but, rather, to find the activations of the model corresponding to the target phone.

We hypothesize that using CTC-based models for GOP would reduce the impact of the alignment errors due to mispronunciations which we mentioned in Sec.\ref{sec:alignment_issues}.
We will show that this method leads to improvements in MDD accuracy not only for the peaky models it was designed for, but for all the other models we have tested.


\subsection{Segmentation-free GOP (GOP-SF)}
\label{sec:gop-segmentation-free}
The second method is segmentation-free (SF) and evaluates the GOP for a particular speech segment without the need for explicitly segmenting the utterance under test.
The motivation for this method is to overcome the following limitations of standard GOP:
1) the evaluation of pronunciation of each phoneme is exclusively based on the corresponding speech segment (see Figure~\ref{fig:gop-goal});
2) the evaluation of GOP is sensitive to the specific alignment ($t_1,t_2$) and does not take into account uncertainty in alignment;
3) the common implementation of GOP uses Viterbi decoding and therefore only considers one path through the model possibly leaving out part of the probability mass; finally,
4) standard GOP does not allow for insertion and deletion errors.

We assume that we have recorded an utterance $O_1^T = \{o_1, \dots, o_T\}$ with a canonical transcription $\Lcano = \{l_1, \dots, l_i, \dots, l_{|\Lcano|}\}$.
We are interested in evaluating the pronunciation of a phone $l_i$.
We can therefore split the canonical transcription into three parts, the left context~(\Lleft) the target phone $l_i$, and the right context~(\Lright):
\begin{equation*}
    \Lcano = \{\underbrace{l_1, \dots, l_{i-1}}_{\Lleft}, \underbrace{l_{i},}_{\text{target phone}} \underbrace{l_{i+1}, \dots, l_{|\Lcano|}}_{\Lright}\}.
    \label{eqn:lcano_definition}
\end{equation*}
As in previous sections, we define $t_1$ and $t_2$ as the start and end frame index for the target phone $l_i$.
\begin{figure}
  \centering
  \includegraphics[width=0.97\columnwidth]{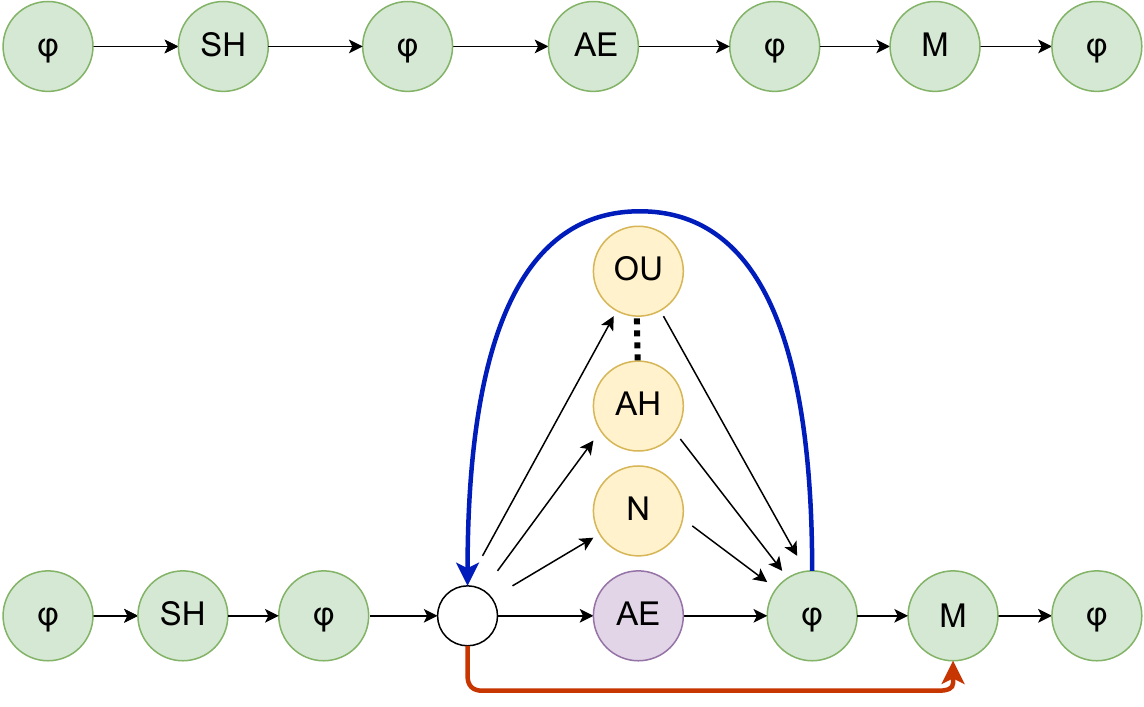} 
      \caption{The figure exemplifies the computation of GOP-SF (Eq.~\ref{eq:gop-sf_implementation}) using a CTC-trained model.
      In the example, $\mathcal{L}_{\text{CTC}}(\Lcano, O_1^T)$ can be computed using the graph at the top, whereas $\mathcal{L}_{\text{CTC}}(\Lsdi, O_1^T)$ using the graph at the bottom.
      For the sake of simplicity we omit the skip connections and self-loop connections.}
  \label{fig:af-gragh}
\end{figure}
Instead of committing to a specific segmentation as in standard GOP, in the proposed segmentation-free GOP (GOP-SF) we compute the log posterior for the target phone $l_i$ given the full observation sequence $O_1^T$ and all the phonemes in the left and right context:
\begin{equation}
    \text{GOP-SF}(l_i) = \log\left(p(l_i|O_1^T, \Lleft, \Lright)\right).
    \label{eq:gop-sf_definition}
\end{equation}
This is the same definition that we introduced in \cite{cao24b_interspeech}, although not explicitly written as in Eq.~\ref{eq:gop-sf_definition}, there.

In this work, we also introduce a version of this definition that is normalized by the estimated length of the model activations for the target speech segment:
\begin{equation}
    \text{GOP-SF-Norm}(l_i) = \frac{\log\left(p(l_i|O_1^T, \Lleft, \Lright)\right)}{\mathbb{E}[t_2-t_1|O_1^T, \Lleft, \Lright]}.
    \label{eq:gop-sf_implementation_normalized}
\end{equation}
The reason for this new definition is to reduce the variance of the GOP estimates with the different lengths of the activations.
This is similar to the standard GOP definition of Eq.~\ref{eqa:gop}, where the posterior is normalized by $|t_2-t_1|$ with the following differences: 1) our normalization factor is not related to the length of the target segment, but, rather, to the length of the activations of the model corresponding to the target segment. This accommodates both peaky and non-peaky models; 2) we do not commit to a specific speech segmentation and incorporate the uncertainty in the alignment into our estimation.

In the following subsections, we give theoretical and practical accounts of how to estimate the numerator and denominator in Eqs.~\ref{eq:gop-sf_definition} and \ref{eq:gop-sf_implementation_normalized}.
\subsection{Segmentation-free target phoneme posterior estimation}
\label{sec:alignment-free_posterior_estimation}
In this section we show how to compute the argument of the log function in Eqs.~\ref{eq:gop-sf_definition} and \ref{eq:gop-sf_implementation_normalized}, that is the alignment-free estimation of the posterior for the target phoneme $p(l_i|O_1^T, \Lleft, \Lright)$.
We first rewrite the expression using the chain rule of probabilities:
\begin{equation}
   p(l_i|O_1^T, \Lleft, \Lright) = \frac{ p(\Lleft, l_i, \Lright|O_1^T)}{p(\Lleft, \Lright|O_1^T)}.
   \label{eq:gop-sf_first_expansion}
\end{equation}

We now consider a specific alignment $(t_1, t_2)$ for the target phone $l_i$ and we define the set of all possible alignments as
\begin{equation}
    \SEG(l_i)=\{(t_1,t_2): i-1 < t_1 \leq t_2 < T-(|\Lcano|-i)\}.
\end{equation}
Because the left context $\Lleft$ and the right context $\Lright$ contain respectively $i-1$ and $|L_C|-i$ symbols, 
the lower bound for $t_1$ and the upper bound for $t_2$ ensure that there is at least one frame for each phone in the left and right context.

We can now expand numerator and denominator in Eq.~\ref{eq:gop-sf_first_expansion} by considering the specific alignment ($t_1, t_2$) and by summing over all possible alignments $\SEG(l_i)$:

\begin{flalign}
    & \frac{p(\Lleft, l_i, \Lright|O_1^T)}{p(\Lleft, \Lright|O_1^T)} = \nonumber \\
    &= \frac{\sum\limits_{(t_1, t_2) \in \SEG(l_i)} p(\Lleft, l_i, \Lright, t_1, t_2|O_1^T)}
    {\sum\limits_{(t_1, t_2) \in \SEG(l_i)} p(\Lleft, \Lright, t_1, t_2|O_1^T)} = \nonumber \\
    &= \frac{\sum\limits_{(t_1, t_2) \in \SEG(l_i)} p(\Lleft, l_i, \Lright|O_1^T, t_1, t_2) p (t_1, t_2|O_1^T)}
    {\sum\limits_{(t_1, t_2) \in \SEG(l_i)} p(\Lleft, \Lright|O_1^T, t_1, t_2) p (t_1, t_2|O_1^T)}.
    \label{eq:gop-sf_second_expansion} 
\end{flalign}

Special attention must be paid to the term $p(t_1, t_2|O_1^T)$ within both sums at the numerator  and denominator  of Eq.~\ref{eq:gop-sf_second_expansion}.
This is the probability of a certain alignment for the $i$th segment, given the observation sequence, but independent of the actual transcription.
We can interpret this as a prior with respect to the transcription over all possible segmentations.
This distribution could be estimated from the training data.
However, in our definition of GOP-SF, we make the simplifying assumption that this distribution is uniform.
Under this assumption $p(t_1, t_2|O_1^T)$ is a constant and can be taken out of the sums and finally cancels out between the numerator and the denominator of Eq.~\ref{eq:gop-sf_second_expansion} which becomes:
\begin{flalign}
    \frac{p(\Lleft, l_i, \Lright|O_1^T)}{p(\Lleft, \Lright|O_1^T)}
    &= \frac{\sum\limits_{(t_1, t_2) \in \SEG(l_i)} p(\Lleft, l_i, \Lright|O_1^T, t_1, t_2)}
    {\sum\limits_{(t_1, t_2) \in \SEG(l_i)} p(\Lleft, \Lright|O_1^T, t_1, t_2) }.
    \label{eq:gop-sf_third_expansion} 
\end{flalign}


Finally, the terms within the sums in Eq.~\ref{eq:gop-sf_third_expansion} can be computed with the CTC's forward variables $\alpha$ and $\beta$ as defined in \cite{CTC}:
\begin{flalign}
    \frac{\sum\limits_{(t_1, t_2) \in \SEG(l_i)} \alpha_{t_1-1}(i-1) \left[\prod_{t=t_1}^{t_2} y_t(l_i)\right] \beta_{t_2+1}(i+1)}
    {\sum\limits_{(t_1, t_2) \in \SEG(l_i)} \alpha_{t_1-1}(i-1) \beta_{t_2+1}(i+1)}.
    \label{eq:gop-sf_forward} 
\end{flalign}

In fact, the numerator in Eq.~\ref{eq:gop-sf_forward} is equivalent to the original definition of the CTC loss for the canonical pronunciation \Lcano and the acoustic features $O_1^T$, except for a $-\log(.)$ term.
Similarly, the denominator can be computed with CTC loss if we define a modified canonical pronunciation $\Lsdi$ as:
\begin{align*}
 \Lsdi &= 
 \{
 \underbrace{l_1, \dots, l_{i-1}}_{\Lleft}, 
 \underbrace{.*,}_{\text{any phoneme sequence}} 
 \underbrace{l_{i+1}, \dots, l_{|\Lcano|}}_{\Lright},
 \}.
\end{align*}
This is the set of all possible transcriptions in which the left and right contexts are equal to the canonical transcription, but we allow any sequence of phonemes in place of the target phoneme $l_i$\footnote{In this and later expressions, we have borrowed notation from regular expressions where ``.'' corresponds to any phoneme, ``*'' is zero or more occurrences, and ``?'' is zero or one occurrence.}.
This corresponds to allowing any number of ``substitution'', ``deletion'' and ``insertion'' errors in pronunciation, justifying the SDI subscript.
This CTC loss can be implemented by defining a graph exemplified in Figure~\ref{fig:af-gragh} (bottom) and running the forward algorithm using this graph.

In summary, combining Eqs~\ref{eq:gop-sf_definition} and \ref{eq:gop-sf_forward}, the GOP-SF can be efficiently computed as the difference between the CTC loss for the pair~(\Lsdi,$O_{1}^{T}$) and the CTC loss for~(\Lcano, $O_{1}^{T}$):
\begin{equation}
    \text{GOP-SF}(l_i) = \mathcal{L}_{CTC}(\Lsdi, O_{1}^{T}) - \mathcal{L}_{CTC}(\Lcano, O_{1}^{T}).
    \label{eq:gop-sf_implementation}
\end{equation}



Note that, because we are considering contributions from all possible segmentations, our method is able to deal with uncertainty in alignment.
Also, the definition of \Lsdi with a graph as in Figure~\ref{fig:af-gragh} gives us flexibility on the kind of pronunciation errors we consider.
If the full graph is used, any substitution (S), deletion (D) and insertion (I) errors are considered.
We refer to this version of the method as GOP-SF-SDI.
If we remove the blue path from the graph, we only consider substitutions and deletions (GOP-SF-SD).
This corresponds to a modified canonical pronunciation
\begin{align*}
 \Lsd &= 
 \{
 \underbrace{l_1, \dots, l_{i-1}}_{\Lleft}, 
 \underbrace{.?,}_{\text{any phoneme or empty}} 
 \underbrace{l_{i+1}, \dots, l_{|\Lcano|}}_{\Lright},
 \}.
\end{align*}

If we also remove the red path, we only consider substitutions as in the traditional GOP (GOP-SF-S) and the modified canonical pronunciation is
\begin{align*}
 L_S &= 
 \{
 \underbrace{l_1, \dots, l_{i-1}}_{\Lleft}, 
 \underbrace{.,}_{\text{any phoneme}} 
 \underbrace{l_{i+1}, \dots, l_{|\Lcano|}}_{\Lright},
 \}.
\end{align*}

In \cite{cao24b_interspeech} we used the unnormalized forward variables $\alpha$s to compute $\mathcal{L}_{CTC}(\Lsdi, O_{1}^{T})$ due of implementation issues.
However, this may lead to numerical underflow errors when multiplying probabilities over longer speech segments.
In this paper, we implement a version of the algorithm that uses the normalized forward variables $\hat{\alpha}_{t}(s)$ defined in \cite{CTC} for both terms in Eq.~\ref{eq:gop-sf_implementation}.
When making comparisons, we refer to this version as GOP-SF-numerical.

In the experiments, we also consider alternative methods to compute the loss, besides standard CTC.

\subsection{Segmentation-free activation length estimation}
\label{sec:estimating_duration}
We now turn to the denominator of Eq.~\ref{eq:gop-sf_implementation_normalized}, that is on the estimation of the model activation length for the target phoneme $\mathbb{E}[t_2-t_1|O_1^T, \Lleft, \Lright]$.


We call $\mathcal{N}_{\text{C}}$ the set of central nodes (yellow and purple) in the graph in Figure~\ref{fig:af-gragh} that correspond to the ``.*'' term in \Lsdi.
Then, the expected value of the duration of the activations corresponding to $l_i$, is the sum over the whole observation sequence of the normalized forward variables $\hat{\alpha}$ corresponding to the nodes in $\mathcal{N}_{\text{C}}$:
\begin{equation}
    \mathbb{E}[t_2-t_1|O_1^T, \Lleft, \Lright] = \sum_{t=1}^{T}\sum_{s\in \mathcal{N}_{\text{C}}} \hat{\alpha}_t(s) := \text{Occ}(i).
    \label{eq:estimate_occupation}
\end{equation}    
This expression can be efficiently computed with the forward algorithm that is also used to estimate the posterior of the target phoneme $l_i$ detailed in the previous section.
Note that the definitions of \Lsdi and \Lsd allow deletion of the target segment as a possible mispronunciation error.
In this case, $\mathbb{E}[t_2-t_1|O_1^T, \Lleft, \Lright]$ will tend to zero and GOP-SF will tend to diverge.
In the actual implementation, therefore, we define a floor value of 1.

We expect this normalization factor to be most relevant for models that are not peaky in time.
Peaky models tend to have activations that span over a single frame, thus making GOP-SF roughly equivalent to GOP-SF-Norm.

\subsection{Computational complexity}
We can compute the complexity of calculating GOP-SF (Eq.~\ref{eq:gop-sf_implementation}) with the help of dynamic programming.
First we note that we only need to evaluate the CTC loss twice, once for the first term $\mathcal{L}(\Lcano, O_1^T)$ and once for the second term $\mathcal{L}(\Lsdi, O_1^T)$ in Eq.~\ref{eq:gop-sf_implementation}.
The graph used to compute $\mathcal{L}(\Lcano, O_1^T)$ (Figure~\ref{fig:af-gragh}, top) contains $|G| = 2|\Lcano|+1$ nodes (the transcription labels interleaved by $\phi$s).
The graph used to compute $\mathcal{L}(\Lsdi, O_1^T)$ (Figure~\ref{fig:af-gragh}, bottom) contains $|G| = 2|\Lcano| + |V|$ nodes, where $V$ is the set of output symbols from the neural network.
The complexity of running dynamic programming on these graphs is therefore $O(T\times |G|) = O(T\times (|\Lcano|+|V|))$, that is, it is linear both in the length of the utterances, in the number of symbols in the canonical transcription augmented by the number of output symbols, which is usually a constant.
For GOP-SF-Norm, the only overhead is the summation in Eq~\ref{eq:estimate_occupation} because the normalized forward variables $\hat{\alpha}$ are already computed.

\subsection{Segmentation-free GOP features}
Similarly to the approaches in~\cite{weigthed-GOP, improved_DNN, WEI2009896}, the GOP-SF of the $i$th phoneme in a canonical sequence can be expanded into a feature vector using LPP~(log posterior probability) and \textbf{LPR}~(log posterior ratio vector):
\begin{equation}
   \label{eqa:af-feature}
   \textbf{FGOP-SF}(l_i) = \{\text{LPP}, \textbf{LPR}(l_i)\} 
\end{equation}
where $\text{LPP} = \log\mathit{p}(\Lcano|O_1^T) = -\mathcal{L}_{\text{CTC}}(\Lcano)$ follows the definition of CTC's log-posterior in Eq.\eqref{eqa:CTC-loss} and \textbf{LPR} is a vector of length $|C|+1$:
\begin{equation}
    \label{eqa:LPP}
    \textbf{LPR}(i) = \left\{\log\left(\frac{\mathit{p}(\Lcano|O_1^T)}{p(L | O_1^T)}\right) \text{ with } L \in \Lsd(l_i) \right\}
\end{equation}
where $\Lsd(l_i)$ has already been defined in Section~\ref{sec:alignment-free_posterior_estimation}.
We do not include insertion errors here because that would result in infinite length for the feature vectors.
Similar to GOP-SF-Norm, we append the expected activation count in Eq.~\ref{eq:estimate_occupation} as an extra dimension of the feature vector which forms:
\begin{equation}
   \label{eqa:af-feature}
   \textbf{FGOP-SF-Norm}(l_i) = \{\text{LPP}, \textbf{LPR}(l_i), \text{Occ}(i)\}. 
\end{equation}

The feature vectors can be fed into multi-dimensional MDD classification models that take the feature vectors as input.

\subsection{Measures of Peakiness}
In Section~\ref{sec:alignment_issues}, we have introduced the peaky behavior of CTC-trained models both with respect to time (POT) and symbols (POS).
Our two methods were introduced in part to cope with this behavior.
Due to the definitions given in Section~\ref{sec:Methods}, we expect the performance of GOP-SA (self-alignment) to increase with both POT and POS, in contrast to standard methods.
We also expect GOP-SF (segmentation-free) to be less affected by POT and POS.
In order to test the effect of POS and POT on the different GOP methods in a reproducible way, 
%
we define objective measures of POS and POT.
 
POT, can be measured by the frequency of activation of the blank symbols~($\phi$).
To estimate this, we use the average of post-softmax activation for $\phi$ over the speech dataset $\mathcal{D}$ for a given model $\mathcal{M}$:
\begin{equation}
    \text{BC}(\mathcal{D},\mathcal{M}) = \frac{1}{N}\sum_{n=1}^N\frac{1}{T_n}\sum_{t \in T_n}p(\phi|O_t^n;\mathcal{M}),
\end{equation}
where the dataset $\mathcal{D}$ contains $N$ utterances, each represented by a sequence of speech features $O^n$ of length $T_n$ and the corresponding transcriptions $L^n$.
We call this the blank coverage (BC).
High BC corresponds to very short activations of the non-blank output symbols and, therefore, to high peakiness in time (POT).

For POS, we borrow the idea from~\cite{CTC-ME} and define the average conditional entropy (ConEn) as:
\begin{equation}
    \text{ConEn}(\mathcal{D},\mathcal{M}) = \frac{1}{N}\sum_{n=1}^N H(p(\pi|L^n,O^n)),
\end{equation}
where
\begin{flalign}
 & H(p(\pi|L^n,O^n)) = \nonumber \\
 &= -\sum_{\pi \in {\{u\in V^{|O|}:\mathcal{B}(u) = L}\}} p(\pi|L^n,O^n)\log p(\pi|L^n,O^n).  
\end{flalign}
Low ConEn corresponds to high peakiness in symbols (POS).
\section{Experimental Settings}
\label{sec:experiments}
\subsection{Data}
We perform our experiments on two datasets including child speech: CMU Kids and speechocean762.

CMU Kids~\cite{CMU_kids} comprises 9.1 hours of speech from children in the range 6--11 years old, and a total of 5180 sentences spoken by 65 females and 24 males. 
In CMU Kids, each utterance is equipped with a phonetic transcription and utterances including mispronunciations are marked.
We determine the distribution of pronunciation errors by comparing\footnote{The comparing procedure is performed only for those utterances that are marked as problematic.
For the correct utterances, on the other hand, we simply label all the canonical phonemes as correct.}
the phonetic transcription with the canonical pronunciation from the CMU pronunciation dictionary provided by Kaldi~\cite{Povey_ASRU2011}.
There are 150899 labeled phonemes in total, among which, 90.2\% are correct; 3.2\% are substitution errors; 2.3\% are deletion errors; and 4.3\% are insertion errors.
We refer to MDD experiments that use the labels obtained this way as ``real errors''.

Following~\cite{cao23_interspeech}, we also create an alternative MDD task based on the CMU~Kids data, where we systematically change each phoneme in the canonical pronunciation to any other phoneme.
We call this task ``simulated errors'' because we pretend that the recorded speech was incorrectly pronounced.

Finally, the speechocean762 dataset~\cite{speechocean762} includes 5000 utterances read by 250 gender-balanced L2 English learners, half of which are children in the range from 5 years- to 15 years-old.
The dataset is collected specifically for pronunciation assessment tasks with rich annotations at different linguistic levels.
We focus on the phoneme-level where each phoneme is labeled as one of the three categories: 0~(mispronounced),
1~(strongly accented) and 2~(correctly spoken).
speechocean762 provides canonical phone-level transcriptions using the same phone inventory as the CMU pronunciation dictionary that we used for CMU Kids.
We were also interested in comparing the results for children and adults.
In order to do this we split the speechocean762 test set according to the reported age of the speaker.
The children (age 6--11) account for 38.4\% of the utterances whereas adults (age 12--43) account for 61.6\%.
Note however, that the results on speechocean762 children are not comparable to those in CMU Kids because the two data sets define different tasks (ternary vs binary assessment).

\subsection{Baseline segmentation model}
The external segmentation used for all the experiments in this study is based on the same baseline segmentation model. 
The model is a context-independent GMM-HMM model that is trained using the standard Kaldi recipe “train-mono.sh” on the 100 hours training set “train-100-clean” from Librispeech~\cite{libri}.
Since the output symbols of the baseline segmentation model match the symbols of the canonical sequences, we used the Kaldi's command ``gmm-align'' for obtaining the segmentation for both CMU Kids and speechocean762 for further experiments.

\subsection{Pre-training and fine-tuning of the acoustic models}
\label{sec:expeirments-AM}
The acoustic models ``DNN'' (CMU Kids baseline) and ``TDNN'' (speechocean762 baseline) are trained from scratch using the Kaldi recipes\footnote{./egs/librispeech/s5/local/\{nnet2/run\_5a\_clean\_100, nnet3/run\_tdnn\}.sh}.
All the other models are based on the wav2vec2.0-large-xlsr-53 \cite{conneau21_interspeech} which is available on Huggingface\footnote{https://huggingface.co/blog/fine-tune-xlsr-wav2vec2} and have a transformer-based structure~\cite{w2v2}.
This model, with the \texttt{feature\_extractor} frozen, was fine-tuned on the Librispeech ``train-100-clean'' set according to the corresponding loss functions:
\begin{itemize}
    \item CE: cross entropy
    \item CTC: connectionist temporal classification
    \item EnCTC: CTC with entropy regularization \cite{CTC-ME}
    \item EsCTC: CTC with equal space constraint \cite{CTC-ME}
\end{itemize}
Fine-tuning was performed for at most 10 epochs or by early stopping based on the validation loss, with learning rate 1e-4 and batch size 32.
The same model was used for the CMU Kids and speechocean762 experiments.


\subsection{Evaluation}
The evaluation on CMU Kids is based on the canonical transcription for each utterance.
Each phone in the canonical transcription is marked as mispronounced if it deviates from the phonetic transcription from the data set.
Then we compute the receiver operative characteristic curve (AUC-ROC) according to various GOP scores, where correctly detecting a mispronunciation is a ``true positive''.
We opt for this threshold-free criterion because we want a robust method to evaluate the performance of our GOP definitions that is not dependent on a specific threshold chosen to perform the MDD task.
This also allows us to use all the data for testing because we do not need to set aside a training set for threshold optimization.
Moreover, AUC-ROC is considered to be robust to label bias which is the case for our data.
We compute AUC-ROC for each phoneme category and report the arithmetic mean.   

For the MDD-oriented dataset speechocean762 the task is to predict three different classes.
In this case, we train and evaluate several typical MDD models using the standard training and test splits.
We follow the recommendations of using Pearson Correlation Coefficient (PCC) from the dataset's paper~\cite{speechocean762} as MDD performance metric.
 
To give a reference on the quality of the different acoustic models in this study, we also report on phoneme recognition results measured with phoneme error rate~(PER). 
Beam-search decoding without language models is used for recognition.

\begin{table}
\centering
\caption{Mispronunciation Detection on CMU Kids}
\label{tbl:cmu_method_comparison}
\begin{tabular}{ lcc }
 \hline\hline
 Method & \multicolumn{2}{c}{AUC (95\% confidence intervals)} \\
 & Simulated errors & Real errors \\
 \hline
 GOP-DNN-Avg~\cite{cao23_interspeech} & 0.824 ($\pm$ 1.6E-3) & 0.796 ($\pm$ 3.8E-3) \\
 GOP-CE-Avg & 0.967 ($\pm$ 7.2E-4) & 0.851 ($\pm$ 3.0E-3) \\
 GOP-CTC-SA & 0.988 ($\pm$ 4.3E-4) & 0.905 ($\pm$ 2.1E-3) \\
 GOP-CTC-SF-S & \textbf{0.989} ($\pm$ 4.1E-4) & 0.891 ($\pm$ 2.4E-3) \\
 GOP-CTC-SF-SD & 0.986 ($\pm$ 4.7E-4) &\textbf{0.914} ($\pm$ 2.0E-3) \\
 GOP-CTC-SF-SDI & 0.938 ($\pm$ 9.9E-4) & 0.859 ($\pm$ 2.9E-3) \\
 \hline\hline
\end{tabular}
\end{table}

\subsection{Experiments}
The set of experiments we include in this paper aims at answering the following questions:

1) \emph{Which GOP definition results in the best MDD performance?}
In this case we test GOP-X-Avg, GOP-X-SA, and the different variants of GOP-X-SF (GOP-X-SF-SDI, GOP-X-SF-SD, GOP-X-SF-S) where ``X'' corresponds to acoustic models trained with different loss functions and architectures as explained in Sec.\ref{sec:expeirments-AM}.



2) \emph{How are the results dependent on the peakiness of the ASR models used for the assessment?}
In this case we test the different GOP definitions in combination with models that are trained with loss functions that are specifically defined to control peakiness: CTC, EnCTC, EsCTC and CE.
We use our best method GOP-SF-SD as segmentation-free version in this experiment and due to numerical problem with CE-trained models, we use the ``numerical'' implementation as discussed in Sec.\ref{sec:gop-segmentation-free}.

3) \emph{How do the results for GOP-SF depend on the length of the context around the target phoneme?}
In this case we create utterances by cropping the original recordings and gradually increasing the context around the target phoneme.
Each time we add one phoneme left and one right and we extract the corresponding utterance with the help of forced alignment.
We go from no context to context length 7 (both right and left) and we compare to the results obtained with the original utterances (full context).

4) \emph{How do our results compare to the state-of-the-art in MDD?}
In order to compare our results to the state-of-the-art, we tested our best method~(GOP-X-SF-SD) on the speechocean762 dataset, and we also include FGOP, that is, segmentation-free GOP features.
Moreover, we compare our new implementations that solve the numerical issues~(denoted as -\textit{numerical}) and the normalized versions of both the scalar GOP and the GOP features. 
Because the focus is on comparing between different GOP and GOP-features definitions, we keep the models for MDD simple in our state-of-the-art comparison.
In particular, we limit our tests to polynomial regression, support vector regression~(SVR) and GOPT as in \cite{GOPT}.
The PCC is computed between the model's output value and the true label.
For the sake of fair comparison, we preserve all the details for evaluations as the baseline papers, e.g, whether to round the output before calculating the PCC; using polynomial order two for polynomial regression; applying the Radial Basis Function~(RBF) kernel for the SVR models etc.
Due to relatively high variance of the GOPT model, same as in~\cite{GOPT}, we run the training of GOPT five times with random initialization, selecting the best model for each run and then averaging the results.
Following the same idea in~\cite{cao23_interspeech} to reduce the variance of the GOPT model during training, we limit the gradients to flow from phoneme-only loss instead of from all multi-aspects losses.

\section{Results}
\label{sec:results}
\subsection{Method comparison}
Table~\ref{tbl:cmu_method_comparison} shows the results comparing the different methods on the CMU Kids data both for simulated and real errors.
AUC values are reported together with 95\% confidence intervals computed according to~\cite{HanleyAndMcNeil1982ROC-CI}.
The methods based on models trained with CTC outperform both methods based on models trained with CE (GOP-DNN-Avg and GOP-CE-Avg).
GOP-CTC-SF-S, that is, the segmentation-free method restricted to substitution errors is significantly better on the simulated errors whereas GOP-CTC-SF-SD, that also allows deletion errors is significantly better on the real errors.
This can be explained by the fact that simulated errors are produced exclusively by substitution, but real errors may include deletions as well.
The standard evaluation criterion used here only considers phones in the canonical transcription.
This means that insertion errors only have an effect on the evaluation if they modify the pronunciation of the neighbouring canonical phones.
This does not allow to show the full potential of GOP-CTC-SF-SDI, that in our tests performs worse than GOP-CTC-SF-SD.
However, defining an ad-hoc evaluation criterion is outside the scope of this study.

\begin{table*}
\centering
\caption{Phone Recognition and Pronunciation Assessment Results vs Peakiness.
}
\label{table:results_vs_peakiness}
\begin{tabular}{ l|ccc|ccc }
 \hline\hline
  & \multicolumn{3}{c|}{Peakiness and Phone Recognition} & \multicolumn{3}{c}{GOP Methods (AUC, 95\% confidence intervals)} \\
 Models & BC (\%) & ConEn & PER (\%) & GOP-X-Avg & GOP-X-SA & GOP-X-SF \\
 \hline
CE         & 42.22 & 22.9886 & 13.25 & 0.860 ($\pm$2.91E-3)~\cite{cao24b_interspeech} & 0.885 ($\pm$2.50E-3) & 0.870 ($\pm$2.75E-3) \\
EnCTC-0.20 & 79.10 & 37.8672 & 11.63 & 0.842 ($\pm$3.19E-3) & 0.860 ($\pm$2.91E-3) & 0.913 ($\pm$2.01E-3) \\
EnCTC-0.15 & 79.55 & 35.9145 & 11.64 & 0.841 ($\pm$3.20E-3) & 0.861 ($\pm$2.81E-3) & \textbf{0.915 ($\pm$1.97E-3)} \\
EnCTC-0.10 & 80.11 & 34.9024 & 11.47 & 0.841 ($\pm$3.20E-3) & 0.864 ($\pm$2.84E-3) &  \textbf{0.914 ($\pm$1.99E-3)} \\
EsCTC      & 88.52 & 14.2370 & 21.06 & 0.580 ($\pm$6.00E-3) & 0.884 ($\pm$2.52E-3) & 0.896 ($\pm$2.31E-3) \\
CTC        & 88.62 & 2.8287  & 11.46 & 0.824 ($\pm$3.46E-3) & 0.909 ($\pm$2.08E-3) &  \textbf{0.914 ($\pm$1.99E-3)} \\
 \hline\hline
\end{tabular}

\end{table*}

\subsection{Performance versus peakiness}
Table~\ref{table:results_vs_peakiness} shows the results of testing the effect of peakiness of the acoustic models on the real errors of the CMU Kids.
All the models were trained up to checkpoint 8000.
The left part of the table reports blank coverage (BC) as a measure of POT and conditional entropy (ConEn) as a measure of POS, as well as phone error rates.
The right part of the table reports MDD results.
The models were ordered in the table in increasing order of peakiness over time (POT).

Standard CTC results in the highest BC and lowest ConEn, which means that CTC is the peakiest model both with respect to POT and POS. 
It is also the best phone recognizer with the lowest PER among the models.

For the EnCTC models, POS decreases as the weight $\beta$ of the entropy term in the loss function increases.
In our tests we varied $\beta$ in the range $[0.10,0.15,0.20]$.
This is not surprising because the entropy term is similar to ConEn.
More interesting is the fact that POT also decreases with $\beta$ as shown by BC.
The EnCTC performance in phone recognition seems to degrade with the peakiness of the models.

The model trained with EsCTC has the highest PER, a similar POT compared to CTC but a much lower POS (higher ConEn).
This is in line with the experiments in~\cite{CTC-ME}, because the model makes the strong assumption that the labels should be uniformly distributed over time which is not always true for speech. 

As we expected, the model trained with CE has the lowest BC because it is trained frame-wise according to baseline segmentation.
Surprisingly, the ConEn of the CE-trained model is still lower than EnCTC implying that the distribution of possible alignment paths are more concentrated for CE-trained model. 



The experimental results for MDD are shown in the right side of Table~\ref{table:results_vs_peakiness} in terms of AUC and 95\% confidence intervals.
When using the standard GOP definition (GOP-X-Avg), the CE model performs the best.
This is expected because GOP-X-Avg uses the same baseline segmentation that was used to train the CE model.
The CTC and EnCTC models are comparable.
We believe that this is because, although the models have different degrees of peakiness, their activation is still within the segments defined by the baseline segmentation.
On the contrary, the AUC of GOP-EsCTC-Avg is particularly low, most probably because the model activations do not necessarily follow standard speech segmentation.
It is therefore possible that GOP is evaluated, in this case, based on activations other than the target phoneme.
This interpretation is also supported by the fact that the use of the self-aligned definition (GOP-EsCTC-SA) gives results that are comparable with the model trained with CE.
However, GOP-X-Avg is clearly outperformed by the proposed GOP-X-SA and GOP-X-SF methods that will be analyzed next.

In general, the proposed GOP-X-SA is always superior to GOP-X-Avg, suggesting that this method is able to focus on the regions corresponding to the relevant model activations.
It is interesting to note that this phenomenon can be observed even for the models trained with CE, for which the mismatch between model activations and baseline segmentation should be minimal.
Among the different models, CTC training performs the best when using GOP-X-SA.
Our interpretation is that the  self-aligned definition of GOP (GOP-X-SA) prefers to work with peaky models~(both POT and POS) such as the model trained with CTC. 
Comparing to the standard definition (GOP-X-Avg), this method makes it possible to take advantage of the highly-discriminative model which is often good for recognition.
This interpretation is also supported by the slight degradation in AUC results compared to standard CTC for the EnCTC models which are less peaky and less performant in terms of PER.
Exceptionally, the behavior of the model trained with EsCTC has by far the worst phone recognition performance.
However, the model's peakiness~(with the second highest BC and second lowest ConEn) could still make it comparable with CE in MDD performance when combined with GOP-X-SA.

Finally, the last column of the table reports the results with the proposed segmentation-free method (GOP-X-SF).
This method obtains the best overall performance for MDD by a significant margin.
Furthermore, according to the confidence intervals it is not possible to determine which acoustic model performs the best with this method because the results obtained with CTC and EnCTC are not statistically different.
This is in accordance with our hypothesis that GOP-X-SF would be less affected by POS of the acoustic models, compared to methods that rely on segmentation.
We can also observe that, by considering all possible segmentations, GOP-X-SF is able to extract much more useful information for MDD not only compared to the external segmentation (GOP-X-Avg) but also compared to an alignment that is matched to the activations of the acoustic model (GOP-X-SA).

\begin{figure*}
    \centering
    \includegraphics[width=\textwidth]{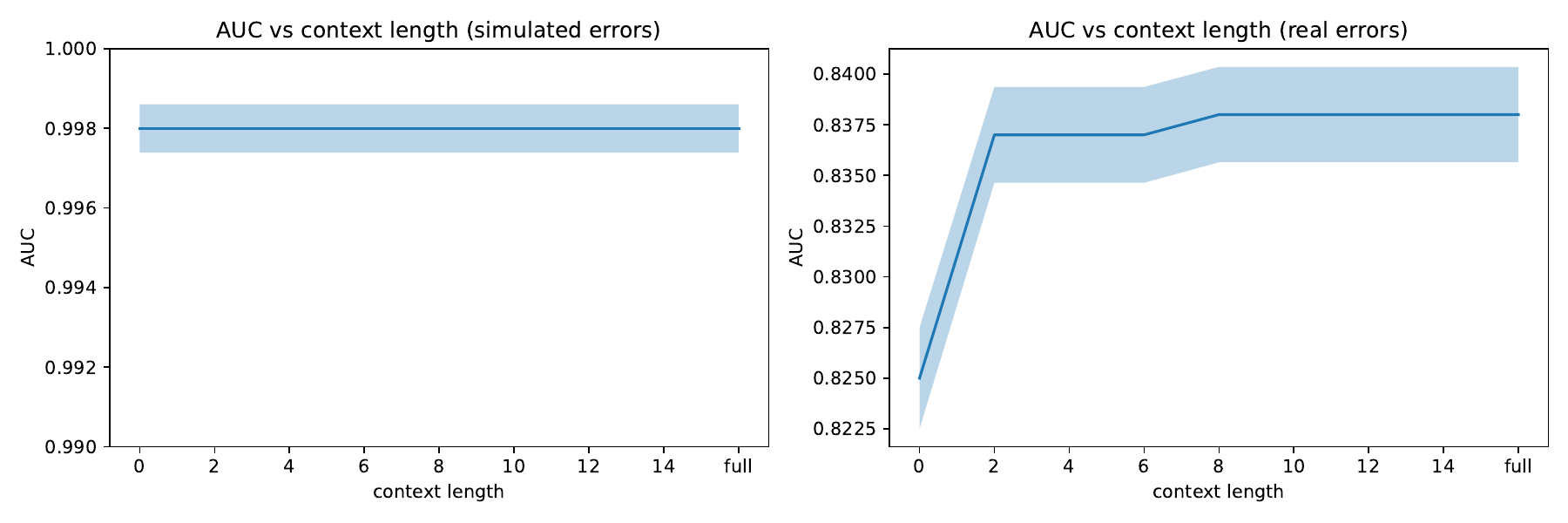}
    \caption{AUC versus context length for simulated and real errors on the CMU Kids data.
    The shaded area shows 95\% confidence intervals computed with \cite{HanleyAndMcNeil1982ROC-CI}.}
    \label{fig:auc_vs_context}
\end{figure*}


\subsection{GOP-SF and context length}
By definition, the segmentation-free GOP method that we have proposed (GOP-SF) is computed considering contributions from the entire utterance, even if those contributions are weighted by how likely it is that each part of the utterance belongs to the target phone.
A reasonable question to ask is how the length of left context~(\Lleft) and right context~(\Lright) affect the pronunciation assessment results.

Figure~\ref{fig:auc_vs_context} displays the AUC results on CMU Kids where we have varied the number of phones in \Lleft and \Lright from 0 to 7 resulting in a total context length from 0 to 14 in increments of 2.
We also report results with the original utterances (full context).
In this case, the length of the context depends on the specific utterance, but it is always greater than~14.
The left plots simulated errors, whereas the right plots real errors.

For simulated errors, the AUC values for different context lengths remain at a high level, and the impact of context length can be neglected.
The results show that GOP-CTC-SF is robust to different context length in the ideal case. 
 
Also for real errors, the dependency on context length is usually under the variability described by the 95\% confidence intervals.
The only exception is when we reduce the context to zero, which corresponds to the self-aligned GOP definition (GOP-SA).
This confirms the superiority of the segmentation-free method that is able to take advantage of the context to assess the pronunciation of the target phoneme.
Using context lengths between 8 and 14 is visually indistinguishable from using the full context length.
This suggests that the information needed to assess the pronunciation of the target phoneme is relatively local in the utterance.

\begin{table}
\centering
\caption{speechocean762, reported with Pearson correlation coefficients}
\label{table:so}
\begin{tabular}{ lcc  }
 \hline\hline
 Features & MDD model & PCC \\
 \hline
 GOP-TDNN~\cite{cao24b_interspeech} & poly. reg. &0.361$\pm$0.008 \\
 GOP-CE-SF-SD & poly. reg. & N/A\\
GOP-CE-SF-SD-\textit{numerical}&poly. reg. & 0.314$\pm$0.008 \\
GOP-CE-SF-SD-Norm& poly. reg. & 0.332$\pm$0.008 \\
GOP-CTC-SF-SD~\cite{cao24b_interspeech}& poly. reg. & 0.433$\pm$0.007 \\
GOP-CTC-SF-SD-\textit{numerical}&poly. reg. & \textbf{0.450}$\pm$\textbf{0.006} \\
GOP-CTC-SF-SD-Norm& poly. reg. & \textbf{0.449}$\pm$\textbf{0.006} \\
 \hline
 FGOP-TDNN~\cite{speechocean762}& SVR &0.441$\pm$0.007 \\
 FGOP-CTC-SF~\cite{cao24b_interspeech}& SVR &0.568$\pm$0.006 \\
FGOP-CTC-SF-\textit{numerical}&SVR&\textbf{0.580$\pm$0.006} \\
FGOP-CTC-SF-Norm&SVR&\textbf{0.581$\pm$0.006} \\
 \hline
  FGOP-TDNN~\cite{GOPT} & GOPT & 0.605$\pm$0.002 \\
  FGOP-CTC-SF~\cite{cao24b_interspeech} & GOPT & 0.618$\pm$0.002 \\
  FGOP-CTC-SF-\textit{numerical} & GOPT & \textbf{0.646}$\pm$\textbf{0.002} \\
  FGOP-CTC-SF-Norm & GOPT & \textbf{0.648}$\pm$\textbf{0.002} \\
     \hline\hline
\end{tabular}

\end{table}

\subsection{Comparison to state-of-the-art}
In order to compare the performance of our best method~(GOP-X-SF-SD) with respect to the state-of-the-art, we report results on the speechocean762 dataset in Table~\ref{table:so}.
The table compares different scalar definitions of GOP with a simple polynomial regression MDD model.
It also includes results with GOP feature vectors in combination with SVR or GOPT MDD models.

Similarly to the previous experiments on CMU Kids, we find that the methods relying on models trained with the CE loss are less performant than those trained with the CTC loss. 
Results for the method ``GOP-CE-SF-SD'' are missing due to numerical problems.
These were solved in the new implementation (``GOP-CE-SF-SD-\textit{numerical}'').
The results are further improved using the definition that normalizes GOP by the estimated length of the target segment (``GOP-CE-SF-SD-Norm'').

In all cases, when combined with a CTC-trained model, our segmentation-free methods (GOP-CTC-SF, and FGOP-CTC-SF) outperform the baseline methods that use traditional GOP with the Time-Delayed-Neural-Network~(TDNN) acoustic models by a significant margin.
We also see improvements when using the implementation which solves the numerical issues compared to our previous results. 
As we expected, the extra normalization steps for the methods are less significant for CTC-trained acoustic models than CE-trained ones. 
The best combination of our method is ``FGOP-CTC-SF-Norm'' which has  $5.01\%$ relatively higher PCC than previously reported. 
We also tested this method on the split of the speechocean762 test set that separates children~(age $\leq 11$) from adults~(age $> 11$).
The corresponding PCCs are $0.635\pm 0.005$ (children) and $0.651\pm 0.004$ (adults).
The slight gap between child and adult result may be reduced by adapting the acoustic models to child speech.

At the time of writing and to the best of our knowledge, the highest phoneme-level PCC obtained for the speechocean762  dataset is $0.693$~\cite{chao23_interspeech}.
The relatively limited PCC improvements of this model with respect to ours are, however, obtained at the cost of higher computational complexity, heavier feature engineering and more sophisticated loss functions that take all the linguistic aspects into account.
For this reason, we believe that our method is still preferable for practical real-world applications.

Finally, we could not find results on pure end-to-end models for MDD on the speechocean762 data.
This is probably due to the limited amount of training data which prevents the proper training of those models.

\section{Conclusions}
\label{sec:conclusions}
In this work, we propose improvements to a framework for pronunciation assessment that we recently introduced in \cite{cao24b_interspeech}.
We propose two methods with the goal of allowing the use of goodness of pronunciation (GOP) features extracted from modern high-performance automatic speech recognition models.
The main idea of our methods is to release the assumption that the GOP evaluation should be performed on a predetermined segment of speech corresponding to the target phone.
We do this either by adjusting the segment under test to the activations of the acoustic models (self-aligned GOP, GOP-SA) or by proposing a definition of GOP that is independent of the segmentation of test utterance (segmentation-free GOP, GOP-SF).

Our theoretical derivations make the underlying assumptions for GOP-SF explicit.
We also provide experimental results exploring a number of properties of our methods.
GOP-SA is always better than traditional GOP regardless of the characteristics of the acoustic model used for the evaluation.
GOP-SF obtains the overall best results for both the CMU Kids and for the speechocean762 material.
On the speechocean762 data we obtain state-of-the-art results keeping the MDD model constant.
We also show how the peakiness of the acoustic models affects the MDD results for the standard GOP definition and for the two proposed methods.
Finally we show that the performance of GOP-SF, that considers the full utterance to assess the pronunciation of the target phone, is not affected by the length of the context.

We believe that the proposed methods are potentially very appealing for phoneme-level pronunciation assessment, both because of their performance but also for their simple implementation and very low computational cost.
\printbibliography

\begin{IEEEbiography}[{\includegraphics[width=1in,height=1.25in,clip,keepaspectratio]{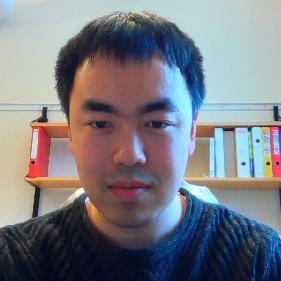}}]{Xinwei Cao}
is a PhD candidate in Signal Processing and Machine Learning at NTNU, Norway, researching speech assessment using state-of-the-art ASR and TTS systems. He previously worked at Cerence Inc., Germany (2019-2022), developing language models for automotive ASR, and at IFlytek Inc., China (2018-2019), building chat-bot systems. He received his M.S. in Computer Science from Karlsruhe Institute of Technology, Germany (2018), and B.S. from Zhejiang University, China (2012). His research interests include speech processing, acoustic and language modeling, neural machine translation and generative AI.
\end{IEEEbiography}

\begin{IEEEbiography}[{\includegraphics[width=1in,height=1.25in,clip,keepaspectratio]{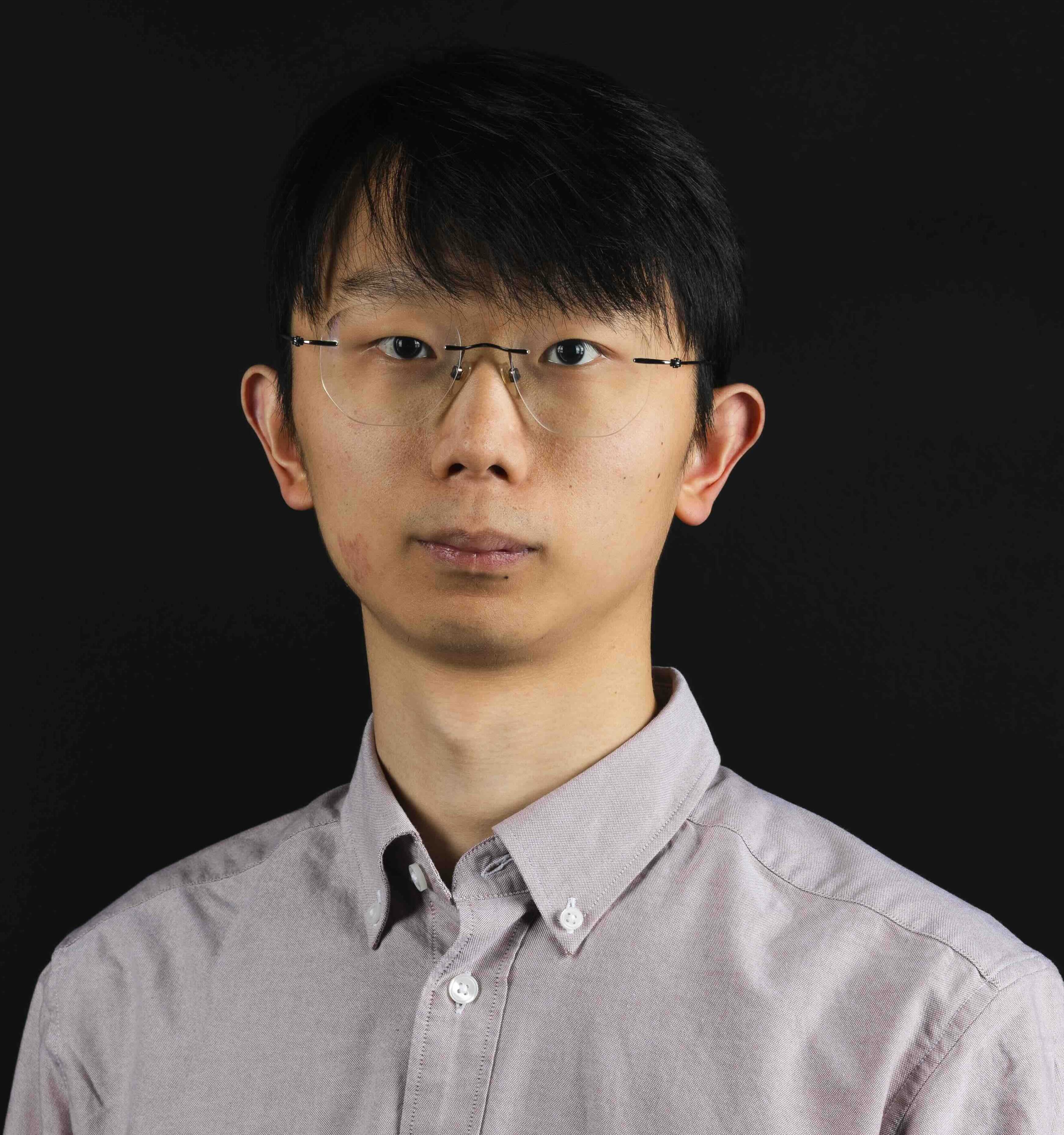}}]{Zijian Fan}
is a PhD candidate in Signal Processing and Machine Learning at NTNU, Norway, researching Child speech ASR sytsems. He received his M.S. in Information and Network Engineering from KTH Royal Institute of Technology, Sweden (2020) and B.S. in Electrical Engineering from Dalian University of Technology, China (2018). His research interests include speech augmentation, end-to-end models, and generative models.
\end{IEEEbiography}

\begin{IEEEbiography}[{\includegraphics[width=1in,height=1.35in,clip,keepaspectratio]{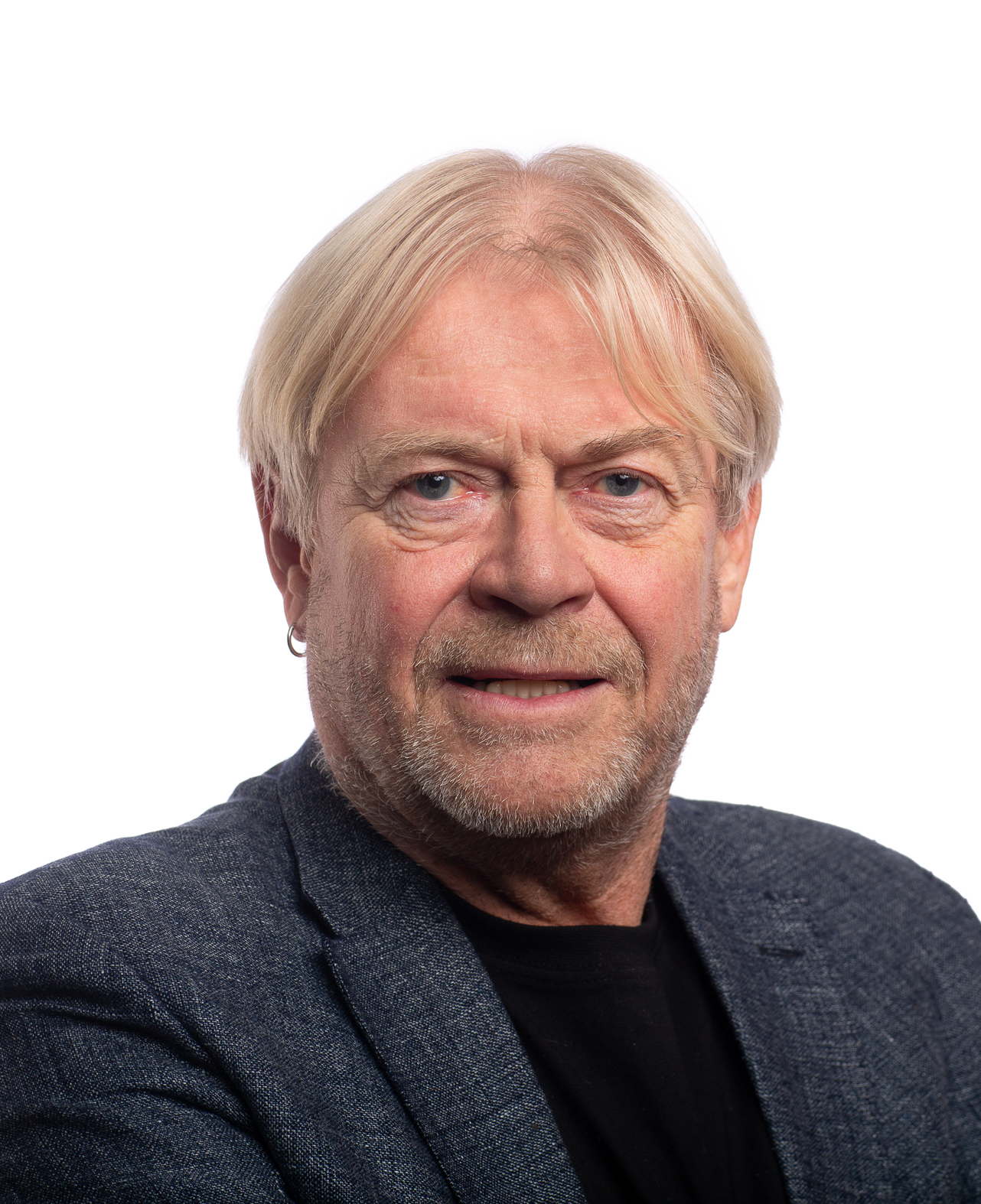}}]{Torbjørn Svendsen}
is a Professor at the Department of Electronic Systems. Professor Svendsen
holds a MScEE, and a PhD both from the NTNU. He is an ISCA Fellow and IEEE Life Senior
Member. He received his dr.ing. (PhD) degree in 1985 from the then Norwegian Institute of
Technology (NTH). He has published over 110 scientific papers and has supervised 18 PhD
graduates and more than 100 Masters students. His research activities have spanned numerous application areas of speech
technology, including speech recognition; speech compression; speaker identification; spoken
language identification and speech synthesis.
He is an ISCA Fellow, and a former Vice
President of ISCA. He is an elected member
to the Norwegian Academy of Technical Sciences.
\end{IEEEbiography}

\begin{IEEEbiography}
[{\includegraphics[width=1in,height=1.25in,clip,keepaspectratio]{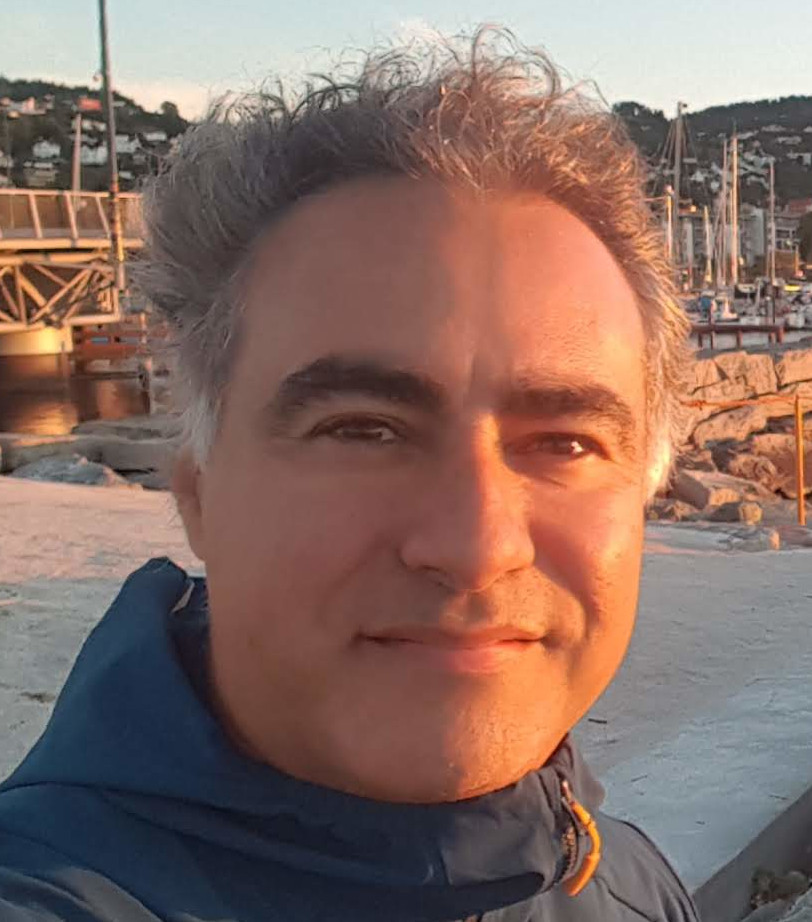}}]{Giampiero Salvi}
(Senior Member, IEEE) is a Full Professor at the Department of Electronic Systems at the Norwegian University of Science and Technology (NTNU), Trondheim, Norway. Prof. Salvi received the MSc degree in Electronic Engineering from Università la Sapienza, Rome, Italy and the PhD degree in Computer Science from KTH. 
He was a post-doctoral fellow at the Institute of Systems and Robotics, Lisbon, Portugal and an Associate Professor at KTH Royal Institute of Technology, Stockholm, Sweden.
He was a co-founder of the company SynFace AB, active between 2006 and 2016.
His main interests are machine learning, speech technology, and cognitive systems and has authored and co-authored more than 110 papers in these areas.
Prof. Salvi is a senior member of IEEE Signal Processing Society (SPS) and a member of the International Speech Communication Association (ISCA).
\end{IEEEbiography}

\end{document}